\documentclass[a4paper,11pt]{article}
\pdfoutput=1 

\usepackage{jheppub} 

\usepackage[T1]{fontenc} 

\usepackage{graphics}
\usepackage{amsmath, amssymb}
\usepackage{multirow}
\usepackage{longtable}
\usepackage{color}
\usepackage{hyperref}
\usepackage{slashed}

\DeclareMathOperator*{\sumint}{%
\mathchoice%
  {\ooalign{$\displaystyle\sum$\cr\hidewidth$\displaystyle\int$\hidewidth\cr}}
  {\ooalign{\raisebox{.14\height}{\scalebox{.7}{$\textstyle\sum$}}\cr\hidewidth$\textstyle\int$\hidewidth\cr}}
  {\ooalign{\raisebox{.2\height}{\scalebox{.6}{$\scriptstyle\sum$}}\cr$\scriptstyle\int$\cr}}
  {\ooalign{\raisebox{.2\height}{\scalebox{.6}{$\scriptstyle\sum$}}\cr$\scriptstyle\int$\cr}}
}

\title{Polarization effects at finite temperature and magnetic field}

\author[1]{Pok Man Lo,\note{Corresponding author.}}
\author{Micha\l {} Szyma\'nski,}
\author{Krzysztof Redlich,}
\author{and Chihiro Sasaki}

\affiliation{Institute of Theoretical Physics, University of Wroclaw,
PL-50204 Wroc\l aw, Poland}

\emailAdd{pokman.lo@uwr.edu.pl}

\abstract{
We study the screening of four-quark interaction by the ring diagram in an effective chiral quark model.  This entails a medium dependent coupling which naturally reduces the chiral transition temperature in a class of models, and is capable of generating an inverse magnetic catalysis at finite temperature.  These results are the first coherent description of inverse magnetic catalysis, anchored to a reliable field-theoretical basis.
    }

\begin{document} 
\maketitle
\flushbottom

\section{Chiral quark model with dressed interaction}

Understanding the properties of QCD matter under extreme conditions, such as
    those created in the laboratory during the ultra-relativistic heavy-ion
    collisions or fill the core of neutron stars, 
    demands a reliable description of 
    how chiral symmetry is restored 
    and how it manifests in a medium of partially deconfined quarks and gluons.  
    Effective models are a robust exploratory tool
    to study a dynamical system.  One of the advantage is the ability to
    temporarily include (or suppress) a certain class of interactions or
    diagrams, thereby examining the effect in isolation.  One can also gain
    insights on the values of phenomenological parameters used and examine
    their connections to the properties of underlying constituents.  
    These make this approach useful to complement the more powerful numerical methods 
    such as lattice QCD (LQCD). 

In this paper we study the in-medium dressing of the four-quark coupling by the
    polarization diagram within an effective chiral quark model.  
    Screening of the potential by ring diagram finds the most famous
    application in regulating
    the long-range Coulomb forces in an electron
    gas~\cite{GellmannBrueckner,ElectronGas}. 
    (See Refs~\cite{Mattuck,Leeuwen_2013} for further discussions 
    in condensed matter theory.)
    When applied to the study of the QCD phase diagram, 
    we shall show that it provides a natural mechanism 
    to reducing the chiral transition temperature in a class of models, 
    and is capable of generating an inverse magnetic catalysis at finite temperature.  

Our starting point is an effective chiral quark model motivated by the Coulomb Gauge QCD~\cite{Govaerts:1983ft,Kocic:1985uq,Hirata:1989qp,Alkofer:1989vr,Schmidt:1995gea,Reinhardt:2017pyr,Quandt:2018bbu,conf}. 
The Lagrangian density reads:
\begin{equation}
    \label{eq:model}
    \begin{split}
    \mathcal{L}(x) &= \bar{\psi}(x) \, (i \slashed{\partial}_x - m) \, \psi(x) \\
    & \quad - \frac{1}{2} \,
    \int d^4 y \,  \rho^a(x) \, V^{ab}(x, y) \, \rho^b(y)
    \end{split}
\end{equation}
where $m$ is the current quark mass, $\rho^a(x) = \bar{\psi}(x) \gamma^0 T^a
\psi(x)$ is the color quark current, and
$T^a$ is a generator of the $SU(N_c)$ symmetry group, with $a = 1, 2, \ldots, N_c^2-1$. 
For the class of models where the interaction potential $V$ is
instantaneous and color-diagonal, i.e.,

\begin{equation}
    V^{ab}(x, y) \rightarrow \delta^{ab} \times \delta(x^0-y^0) \,
    V(\vec{x}-\vec{y}),
\end{equation}
the gap equation for the dynamical quarks has been 
derived~\cite{conf}. 
It can be summarized as follows:

\begin{equation}
        \label{eq:sinv}
        S^{-1}(p) = \slashed{p} - m - \Sigma(p)
\end{equation}
where
\begin{equation}
        \label{eq:sfquark}
    \Sigma(p) = C_F \, \int \frac{d^4 q}{(2 \pi)^4} \, V(\vec{p}-\vec{q}) \, i \, \gamma^0 S(q) \gamma^0.
\end{equation}
The constant $C_F = \frac{N_c^2-1}{2 N_c}$ is introduced via 
the quadratic Casimir operator. 
Note that the use of the dressed quark propagator $S(q)$ within the
integral in Eq.~\eqref{eq:sfquark} means that 
the gap equation is to be solved self-consistently.
A particularly transparent case to consider is a contact interaction, where we replace

\begin{equation}
V(\vec{p}-\vec{q}) \rightarrow V_0.
\end{equation}
The solution to the gap equation~\eqref{eq:sinv} becomes

\begin{equation}
    \label{eq:cgap}
    \begin{split}
        M &= m + {\rm Tr} \,(\Sigma) / {\rm Tr} \, (I) \\
        &= m + C_F \, V_0 \, \int \frac{d^3 q}{(2 \pi)^3} \,
        \frac{M}{2 E} \\
        &\quad \times (1 - 2 \, N_{\rm th}(E)),
    \end{split}
\end{equation}
where $E = \sqrt{q^2 + M^2}$ and $N_{\rm th}(E) = ({e^{\beta E}+1})^{-1} $ 
is the Fermi-Dirac distribution, with temperature $T = 1/\beta$.
Equation~\eqref{eq:cgap} is of the same form as the familiar result for quark mass
($N_f$ flavors) in the model of Nambu-Jona and Lasinio
(NJL)~\cite{Klevansky1992}. The latter has a Lagrangian density

\begin{equation}
    \begin{split}
    \mathcal{L}_{\rm NJL}(x) &= \bar{\psi}(x) \, (i \slashed{\partial}_x - m) \, \psi(x) \\
    & \quad + G_{\rm NJL} \, \left( (\bar{\psi}(x) \, {\psi}(x))^2 + 
        (\bar{\psi}(x) i \gamma_5 \vec{\tau} \psi(x))^2 \right).
    \end{split}
\end{equation}
To leading order the gap equations of the two models would agree 
if we identify 

\begin{equation}
      \label{eq:match}
      C_F \, V_0 \leftrightarrow 4 \, N_c \, N_f \, (2 \, G_{\rm NJL}).
\end{equation}
There are, however, important differences in their theoretical origin:
In the current model, the quark self energy in Eq.~\eqref{eq:sfquark} 
comes from the Fock (exchange) diagram, 
rather than the Hartree diagram in an NJL model.
In fact, the Hartree term vanishes in the current model~\eqref{eq:model} 
due to ${\rm Tr} \, T^a = 0$.
This gives the characteristic color factor in Eq.~\eqref{eq:match}, 
instead of a simple scaling by $N_c$.
Furthermore the quark current here, as dictated by the Coulomb gauge, 
is of (Dirac and color) vector nature, 
rather than a scalar current in the NJL case.
This allows us to formally identify $V_0$ 
with the longitudinal gluon propagator, 
and select the relevant Dirac structure 
of the polarization tensor when dressing the potential.

\section{Polarization tensors}

The in-medium dressing of the interaction potential $V_0$ 
by the polarization tensor $\Pi_{00}$ is implemented via~\cite{conf}

\begin{equation}
    \label{eq:dress}
    \tilde{V_0}^{-1} = {V_0}^{-1} - \frac{1}{2} \, N_f \, \Pi_{00}
\end{equation}
where the polarization is further approximated by the ring diagram
\begin{equation}
    \label{eq:pi00}
    \begin{split}
        \Pi_{00}(p^0, \vec{p}) &= \frac{1}{\beta} \, \sumint {\rm Tr} \left( \gamma^0
        S(q) \gamma^0 S(q+p) \right).
    \end{split}
\end{equation}
Here $\sumint$ denotes a Matsubara sum over the fermionic frequencies
($\omega_n = (2n+1) \, \pi/\beta$), and an integral over the momenta $d^3 q$.
Equation~\eqref{eq:dress} describes the screening of the gluon propagator by the Debye mass. The factor of $\frac{1}{2}$ in Eq.~\eqref{eq:dress} comes from the color structure, i.e. ${\rm Tr} \, T^a T^b = \frac{1}{2} \, \delta^{ab}$, and is essential to reproduce the known result~\cite{Kapusta:2006pm} of the perturbative Debye mass for QCD, rather than for QED.

To simplify the discussion, we shall consider only the screening by the static, finite temperature part of $\Pi_{00}$.
(The static, vacuum part of $\Pi_{00}$ naturally vanishes.)
One way to obtain the quantity is by a direct numerical evaluation of Eq.~\eqref{eq:pi00}. 
A more intuitive way to obtain the result is via 
a formal relation to the thermal pressure of a free (single species) fermion gas at finite
temperature and vanishing chemical potential:

\begin{equation}
    \label{eq:kapusta2}
    \begin{split}
        \Pi_{00}(p^0=0, \vec{p}\rightarrow \vec{0}) &= \frac{1}{\beta} \, \sumint {\rm Tr} \left( \gamma^0
        S(q) \gamma^0 S(q) \right) \\
        &= -\frac{\partial}{\partial \mu} \, \frac{1}{\beta} \, \sumint {\rm Tr} \left( \gamma^0
        S \right) \\
        &= -\frac{\partial^2}{\partial \mu \partial \mu} \, \frac{1}{\beta} \, \sumint {\rm
        Tr} \, \ln S^{-1}.
    \end{split}
\end{equation}
Here $S^{-1}(q) = (i \, \omega_n + \mu) \, \gamma^0 - \vec{q} \cdot \vec{\gamma} - M$.
It is particularly useful to realize that 

\begin{equation}
    \label{eq:1step}
    \begin{split}
        \frac{1}{\beta} \, \sumint {\rm Tr} \left( \gamma^0
        S \right) &= \int \frac{d^3 q}{(2 \pi)^3} \, 2 \times \\
        &\quad \left(
        \frac{1}{e^{\beta ( E-\mu )}+1} - \frac{1}{e^{\beta ( E+\mu )}+1} \right).
    \end{split}
\end{equation}
Taking one further $\mu$-derivative and setting $\mu \rightarrow 0$ at the end, 
yields an explicit expression of the electric mass:~\footnote{
    The perturbative Debye mass is $m_{el}^2 = g^2 \times \hat{m}_{el}^2$.
    }

\begin{figure}
	\resizebox{0.85\textwidth}{!}{%
	\includegraphics{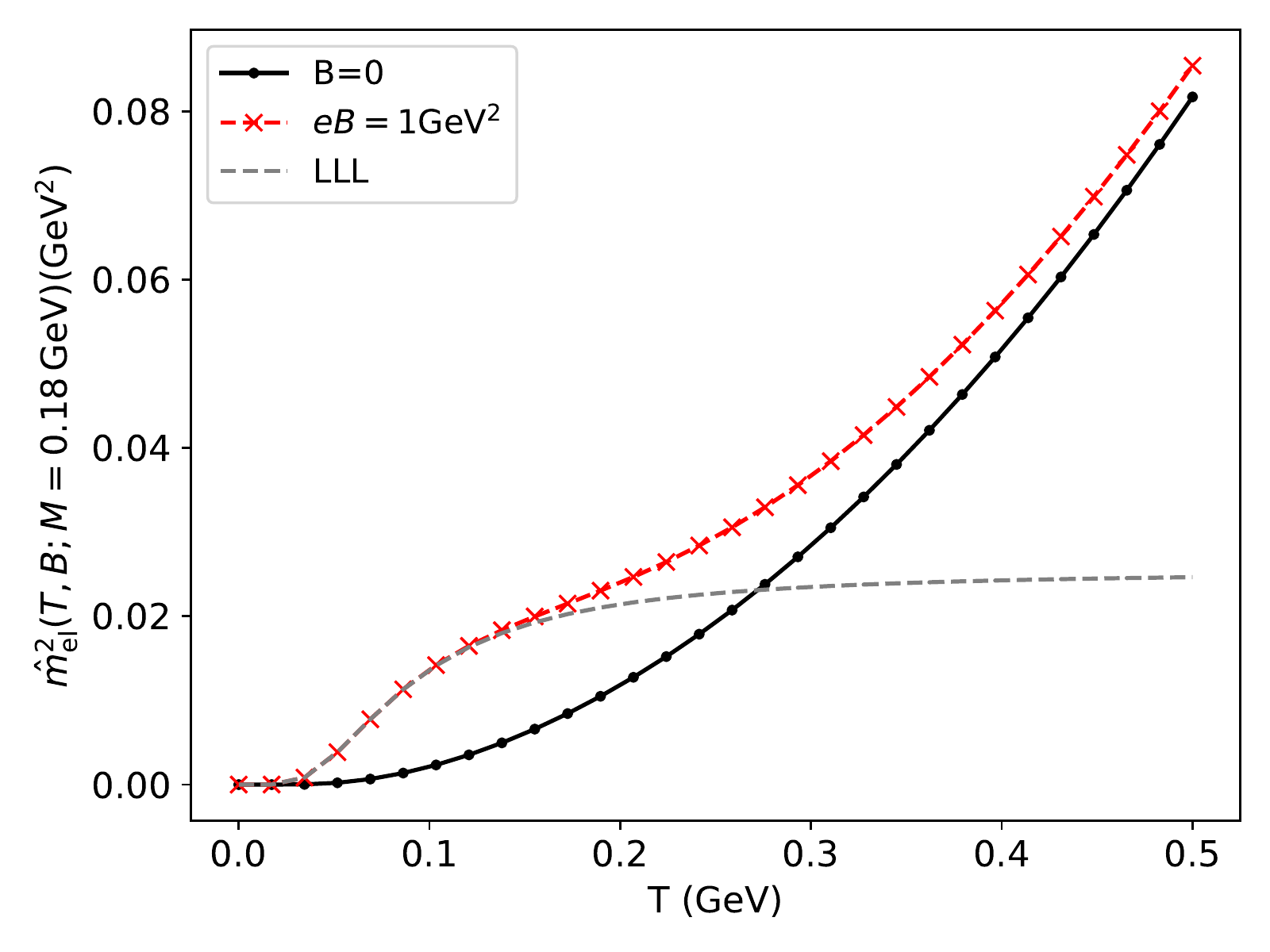}
	}
        \caption{The electric mass (Eq.~\eqref{eq:electric_mass}) for the 2-flavor case at fixed quark mass $M=0.18$ GeV, 
        as a function of temperature. 
        The results in a finite magnetic field are calculated using Eq.~\eqref{eq:electric_massB}
        with physical electric charges of $(u, d)$ quarks.}
\label{fig1}
\end{figure}

\begin{equation}
    \label{eq:electric_mass}
    \begin{split}
    \hat{m}_{el}^2 &= -\frac{1}{2} \, N_f \,  \Pi_{00}(p^0=0, \vec{p} \rightarrow \vec{0}) \\
        &= \frac{1}{2} \, N_f \times  \int \frac{d^3 q}{(2 \pi)^3} \, 4 \beta N_{\rm th} (1-N_{\rm th}).
    \end{split}
\end{equation}
In particular, we note the following interesting limits:
(1) at $M \rightarrow 0$ (or large $T$),
            \label{eq:lim1}
        \begin{equation}
            \hat{m}_{el}^2 \approx \frac{1}{2} \, N_f \times \frac{T^2}{3};
        \end{equation}
and (2) at large $M$ (or small $T$), where the Boltzmann approximation is
valid,
        \begin{equation}
            \label{eq:lim2}
            \hat{m}_{el}^2 \approx \frac{1}{2} \, N_f \times \frac{2}{\pi^2} \, M^2 \,
            ( K_2(M/T)-K_2(2M/T) ),
        \end{equation}
        where $K_2$ is the modified spherical Bessel function of the second kind.
In Fig.~\ref{fig1} we demonstrate a numerical calculation of the electric mass
Eq.~\eqref{eq:electric_mass} at fixed value of $M=0.18$ GeV (a typical value
near the transition), and $N_f=2$ flavors. 
The analytic limits in Eqs.~\eqref{eq:lim1} and \eqref{eq:lim2} 
can be readily verified.~\footnote{Including the second order term of the Boltzmann approximation in Eq.~\eqref{eq:lim2} 
gives a good estimate even at high temperature, i.e. $3/\pi^2 \times T^2 \approx
0.304 \, T^2$, instead of the full quantum result $1/3 \times T^2$.}
As preparation for the analysis of the inverse magnetic catalysis at finite
temperature~\cite{Andersen:2014xxa,Miransky:2015ava}, 
we extend the electric mass 
in Eq.~\eqref{eq:electric_mass} (per flavor) to that with a finite magnetic
field:

\begin{equation}
    \label{eq:electric_massB}
    \begin{split}
        \hat{m}_{el}^2 &= \frac{1}{2} \,  \frac{\vert e_f \vert B}{2 \pi} \,
        \sum_{n=0}^\infty \, \frac{1}{2} \, \alpha_n \times \\
        &\quad \int \, \frac{d q_z}{2 \pi} \, 4 \beta N_{\rm th}(E_{q_z, n}) (1-N_{\rm th}(E_{q_z, n})),
    \end{split}
\end{equation}
where $\alpha_n = 2 - \delta_{n0}$ and $e_f$ is the electric charge of the species.
Note that the effective dispersion relation in the presence of magnetic field
$B$ becomes:

\begin{equation}
    E_{q_z, n} = \sqrt{q_z^2 + 2 \, n \times \vert e_f \vert  B + M^2}.
\end{equation}
Examining in particular the contribution from the lowest Landau level (LLL) ($n=0$),
we get

\begin{equation}
    \hat{m}_{el}^2 \approx \frac{1}{2} \,  \frac{\vert e_f \vert B}{4 \pi} \, \int
        \, \frac{d q_z}{2 \pi} \, \frac{4 \beta e^{\beta \sqrt{q_z^2+M^2}}}{(e^{\beta \sqrt{q_z^2+M^2}}+1)^2 },
\end{equation}
which for massless quarks reduces to~\footnote{
This gives an alternative derivation of the result in Ref.~\cite{debye}.
}

\begin{equation}
    \label{eq:LLL}
    \hat{m}_{el}^2 \rightarrow \frac{1}{2} \,  \frac{\vert e_f \vert B}{2 \pi^2}.
\end{equation}
The limit is realized by the LLL line in Fig.~\ref{fig1} at high temperature.
Note in particular that the full result follows the LLL at low temperature, and eventually reaches the limit~\eqref{eq:lim1} at high temperature.

\begin{figure*}
	\resizebox{\textwidth}{!}{
	\includegraphics{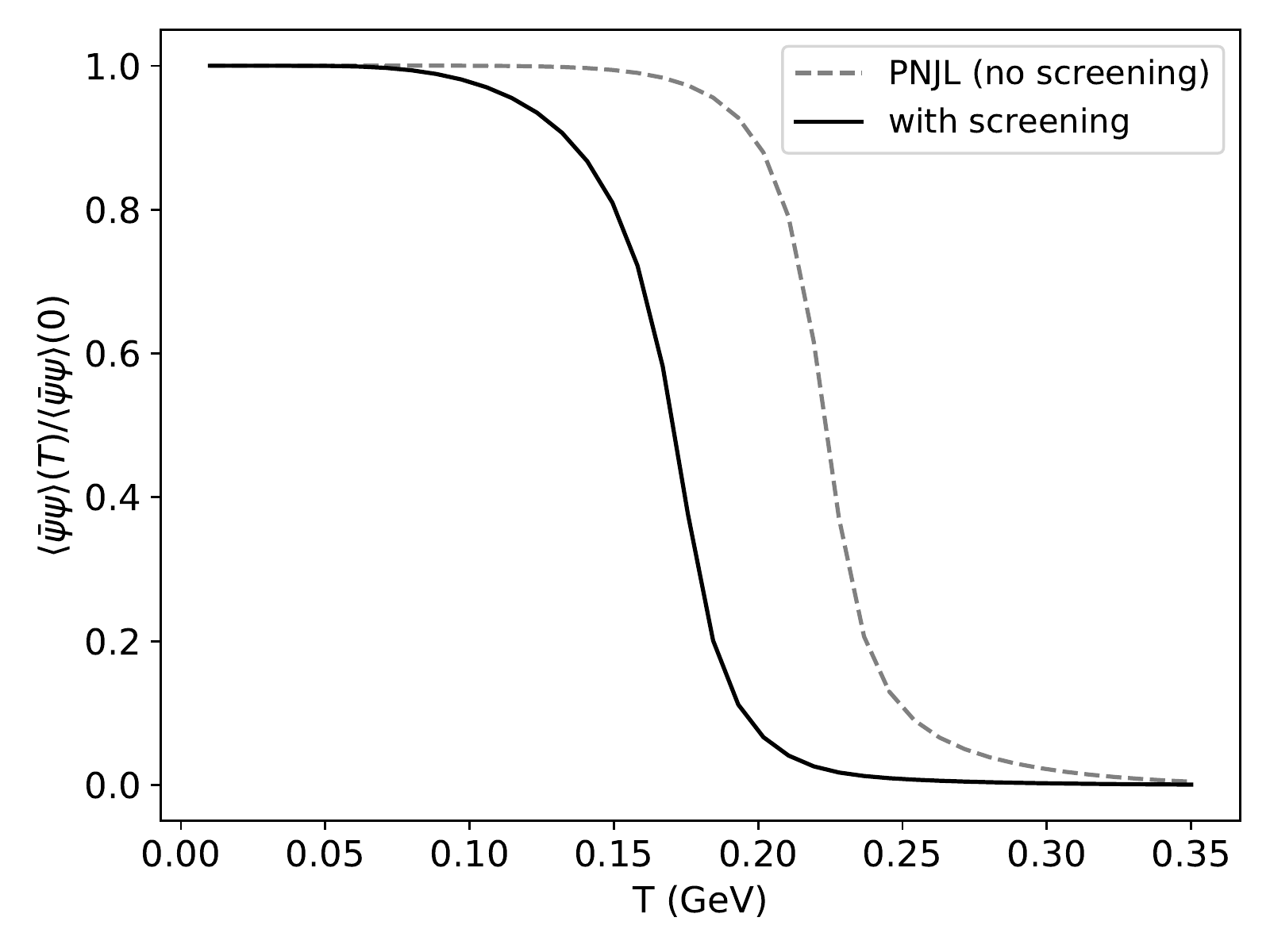}
	\includegraphics{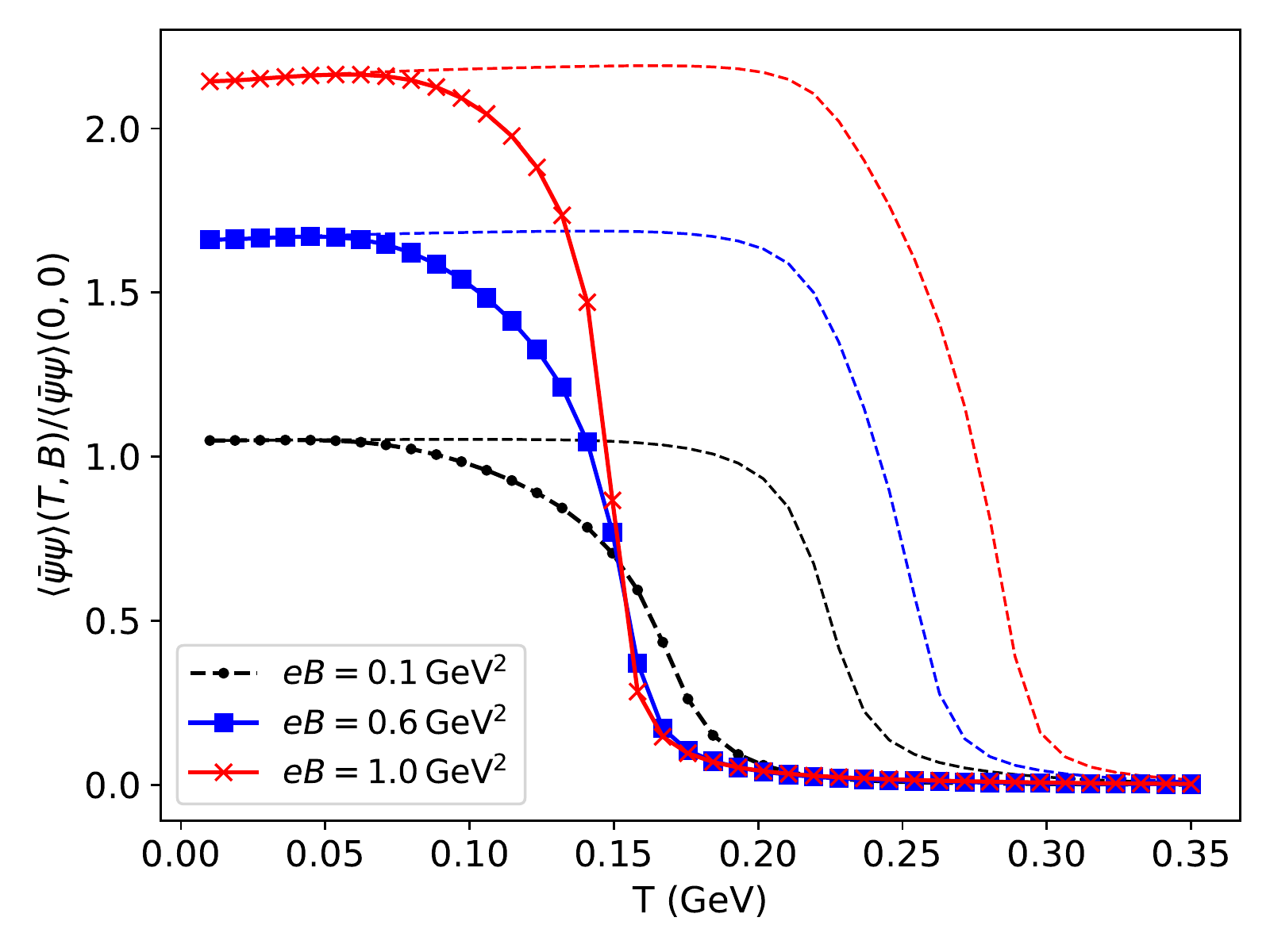}
	}
        \caption{The chiral condensate (with current quark mass contribution
        subtracted), normalized to the vacuum value, versus the temperature, at zero (left) and finite (right) magnetic field. 
        Dashed lines represent results obtained from a PNJL model without 
        screening effect. 
        The model with dressed coupling is capable of producing the inverse magnetic catalysis at finite temperature.}
\label{fig2}
\end{figure*}

\section{Results}

\subsection{zero magnetic field}

A standard NJL model lacks a description of confinement. 
This is usually remedied 
by considering an additional coupling of quarks to the Polyakov loop $\ell$: 
a Polyakov-NJL (PNJL) model~\cite{Fukushima:2003fw,pnjl,Sasaki:2006ww,Fukushima:2017csk}.
One known issue of this class of models is that it 
overestimates the chiral transition temperature: 
typically giving $T_{\rm ch} \approx 220 $ MeV instead of 
the LQCD result $T_{\rm ch} \approx 156.5$ MeV~\cite{Td1,Td2}.
A common way to deal with this problem is by an {\it ad hoc} rescaling of the $T_d$
parameter, i.e. the deconfinement transition temperature in a pure gauge
theory, in the pure gauge potential $U_{\rm glue}$: 
from the physical value of $T_d \approx 270$ MeV, 
to a lower value at $T_d \approx 200$ MeV~\cite{shift1,shift2,mediumG1}. 
This is far from ideal as it points to some missing interactions in the original model. 
Here we propose the screening of the interaction by polarization 
as a natural resolution to the problem.
For this purpose, 
we generalize our model to include the coupling to $\ell$ 
by implementing a statistical confinement scheme,
i.e., by replacing the thermal weight $N_{\rm th}(E)$ in Eq.~\eqref{eq:cgap} with
($N_c=3$)~\cite{Skokov:2010wb,Skokov:2010uh,su3pot} 

\begin{equation}
    \label{eq:weight}
    \begin{split}
N_{\rm th}(E, \ell) &\rightarrow \frac{1}{3} \, \sum_{j=1}^3
\frac{\hat{\ell}_F^{(j)}}{e^{\beta E}+\hat{\ell}_F^{(j)}} \\
        &= \frac{1}{3} \, \frac{3\ell \, e^{-\beta E} + 6 \ell \, e^{-2 \beta
        E} + 3 e^{-3\beta E}}{1 + 3 \ell \, e^{-\beta E} + 3 \ell \, e^{-2\beta
        E} + e^{-3\beta E}},
    \end{split}
\end{equation}
where $\hat{\ell}_F^{(j)}$ are entries of the Polyakov loop operator in the fundamental representation
\begin{equation}
    \hat{\ell}_F = {\rm diag} \, \left( e^{i \gamma_1}, 1, e^{-i
    \gamma_1} \right),
\end{equation}
related to the Polyakov loop $\ell$ via

\begin{equation}
    \ell = \frac{1}{3} \, {\rm Tr} \, \hat{\ell}_F = \frac{1}{3} \, (1 + 2 \cos \gamma_1).
\end{equation}
The final set of gap equations for quarks becomes

\begin{equation}
    \label{eq:newgap1a}
        M = m + C_F \, \tilde{V}_0 \, M \times I_0(T; M, \ell),
\end{equation}
where

\begin{equation}
    \begin{split}
        I_0 &= I_0^{\rm vac} + I_0^T, \\
        I_0^{\rm vac}  &= \int \frac{d^4 q_E}{(2 \pi)^4} \, \frac{1}{q_E^2 + E^2} \, \mathcal{R}_{4D}(q), \\
        I_0^T &= -\int \frac{d^3 q}{(2 \pi)^3} \, \frac{1}{2 E}
        \times 2 N_{\rm th}(E, \ell).
    \end{split}
\end{equation}
The vacuum piece $I_0^{\rm vac}$ requires regularization, e.g., by a 4D regulator
$\mathcal{R}_{\rm 4D}(q)$, 
while the finite temperature piece $I_0^T$ requires no regularization.~\footnote{In computing $I_0$ under a general regularization scheme, 
the finite temperature piece can be extracted via 
\begin{equation*}
    I_0^T = \lim_{\Lambda \rightarrow \infty} \,
    \left(I_0(T, \Lambda) - I_0(T \rightarrow 0, \Lambda)
    \right).
\end{equation*}
This serves as an alternative way to define the finite temperature contribution in a
regularization independent manner~\cite{regulators,Lo:2013lca}.}
The screened potential is obtained from

\begin{equation}
    \label{eq:newgap1b}
    \tilde{V}_0(T; M, \ell) = \frac{1} {V_0^{-1} +
        \hat{m}_{el}^2 (T; M, \ell) }.
\end{equation}
Based on Eq.~\eqref{eq:1step} and the statistical weight~\eqref{eq:weight}, 
we arrive at the thermal factor appropriate for the calculation of electric
mass in Eq.~\eqref{eq:electric_mass}:

\begin{equation}
N_{\rm th} \,  (1- N_{\rm th}) \rightarrow \frac{1}{3} \, \sum_{j=1}^3
    \frac{e^{\beta E} \, \hat{\ell}_F^{(j)}}{(e^{\beta
    E}+\hat{\ell}_F^{(j)})^2}.
\end{equation}
The expectation value of the Polyakov loop needs to be determined from another
gap equation,

\begin{equation}
    \label{eq:newgap2}
    \frac{\partial}{\partial \ell} \, (U_{\rm glue}(\ell) + U_Q(M, \ell)) = 0,
\end{equation}
for a given $U_{\rm glue}(\ell)$ and the quark potential
$U_{Q}(M, \ell)$, the latter describes the coupling of the Polyakov loop with quarks. 
These potentials have been studied 
extensively~\cite{Fukushima:2003fw,Sasaki:2006ww,Lo:2013hla} and will not be repeated here.
In this work we employ the pure gauge potential in Ref.~\cite{Lo:2013hla}.
This potential is unique in the sense that the locations of minima 
and the curvatures around them are constrained with the SU(3) pure gauge LQCD data~\cite{sus}.

The screening of $V_0$ in Eq.~\eqref{eq:newgap1b} introduces further field dependence in $M$ and $\ell$, which, 
depending on the model parameters chosen, can induce a first order phase
transition instead of a crossover~\cite{conf,Lo:2013lca}. 
This is an interesting point and we shall defer its detailed study to a
separate paper.
As a first numerical demonstration of the effect of polarization, 
we consider a further approximation of $ \tilde{V}_0(T; M, \ell) \approx
\tilde{V}_0(T; \langle M \rangle, \ell)$, and fix $\langle M \rangle = 0.136$ GeV.
The set of model parameters is given by: 
$\Lambda = 1.076$ GeV, $G_{\rm NJL} \, \Lambda^2 = 4.232$, $\mathcal{R}_{4D}(q)
= e^{-q^8/\Lambda^8}$, 
and a current quark mass $m = 5$ MeV. 
In the vacuum, this gives: $f_\pi = 92.9$ MeV, $m_\pi = 137.8$ MeV, and
$\langle \bar{\psi} \psi \rangle = -(250 \, {\rm MeV})^3$ (per flavor).
Solving Eqs.~\eqref{eq:newgap1a}, ~\eqref{eq:newgap1b} and ~\eqref{eq:newgap2} consistently, 
we obtain the results in Fig.~\ref{fig2} (left).
The screening of the effective four-quark coupling by the
electric mass leads to a substantial drop in the chiral transition temperature.

\subsection{finite magnetic field}

While most effective chiral models can capture 
the magnetic catalysis in vacuum, 
i.e. chiral condensate increases in magnitude at $T=0$ on increasing $B$,
they predict the opposite trend from LQCD calculations~\cite{Bali:2011qj,Bruckmann:2013oba}
on the magnetic field dependence of the 
chiral transition temperature~\cite{Boomsma:2009yk,Lo:2020ptj}.
It is therefore of interest to explore whether 
introducing the in-medium coupling via Eq.~\eqref{eq:dress} 
could accommodate a decreasing transition temperature with $B$.
With the same model setup we perform a numerical study of the chiral condensate, 
now at finite magnetic field $B$. 
The result is shown in Fig.~\ref{fig2} (right). 
Without the polarization screening in Eq.~\eqref{eq:dress},
the PNJL model predicts an increasing chiral transition temperature with $B$, 
evident from the fact that 
the condensate remains large for a wider range of temperatures.
The dressed interaction is capable of reversing this trend.
Note that the electric mass increases substantially at intermediate temperature 
with increasing $B$, which can drive the phenomenon of inverse magnetic catalysis.
The need for a $B$-dependent coupling has been motivated 
in Refs.~\cite{mediumG1,mediumG2}, 
and Eq.~\eqref{eq:dress} is a theoretical justification 
for introducing such a medium dependence.

\section{Conclusion}

The promising results motivate further research 
to explore how polarization effects modify the chiral quarks 
at finite baryon density.
A systematic examination of the many consequences of 
chiral symmetry restoration is mandatory to 
construct more stringent predictions to probe possible signals of criticality
and to guide the search of the enigmatic critical endpoint 
by the heavy-ion collision programme.
To make progress it is essential to 
incorporate other important physical effects absent in the current model.
Chief among them is to restore the momentum dependence of 
various quark dressing functions, i.e. constituent mass function and 
the wavefunction renormalization, 
based on better modeling of the confining potential~\cite{conf}. 
The external magnetic field thus provides an additional handle to 
probe the relations between the quark propagator and the gluon potential.
Another important aspect is the explicit treatment of dynamical gluons (and ghosts) 
in the confinement model~\cite{eric_cgauge,Lo:2020ptj}. 
Unlike the QED case, the electric mass of gluons contains a pure glue
contribution, which will further modify the in-medium four-quark coupling.
Due to its connection to the Cartan curvature masses~\cite{Lo:2020ptj}, 
this needs to be determined consistently with the Polyakov loop potential.
~\footnote{Beside the statistical confinement models discussed here, 
there are also confinement models which suppress the quark energy spectrum 
via an infrared divergent mass (and wave function
renormalization)~\cite{conf}.} 
Finally, 
a lattice computation of the gluon Debye mass at low temperature (with a finite
magnetic field) 
could validate the relevance of ring diagram screening to inverse magnetic catalysis.

\acknowledgments

This study receives supports from the Polish National Science Center (NCN) under the Opus grant no. 2018/31/B/ST2/01663.
M.S. acknowledges the support from Preludium grant no. 2020/37/N/ST2/00367.
K.R. also acknowledges the support by the 
Polish Ministry of Science and Higher Education.



\bibliographystyle{jhep}
\bibliography{ref}

\providecommand{\href}[2]{#2}\begingroup\raggedright\begin{thebibliography}{10}

\bibitem{GellmannBrueckner}
M.~Gell-Mann and K.A.~Brueckner, \emph{Correlation energy of an electron gas at
  high density}, \href{https://doi.org/10.1103/PhysRev.106.364}{\emph{Phys.
  Rev.} {\bfseries 106} (1957) 364}.

\bibitem{ElectronGas}
D.~Bohm and D.~Pines, \emph{A collective description of electron interactions:
  Iii. coulomb interactions in a degenerate electron gas},
  \href{https://doi.org/10.1103/PhysRev.92.609}{\emph{Phys. Rev.} {\bfseries
  92} (1953) 609}.

\bibitem{Mattuck}
R.D.~Mattuck, \emph{{A Guide to Feynman Diagrams in the Many Body Problem
  (Second Edition)}} (1976).

\bibitem{Leeuwen_2013}
G.~Stefanucci and R.~van Leeuwen, \emph{Nonequilibrium Many-Body Theory of
  Quantum Systems: A Modern Introduction}, Cambridge University Press (2013),
  \href{https://doi.org/10.1017/CBO9781139023979}{10.1017/CBO9781139023979}.

\bibitem{Govaerts:1983ft}
J.~Govaerts, J.E.~Mandula and J.~Weyers, \emph{{A Model for Chiral Symmetry
  Breaking in {QCD}}},
  \href{https://doi.org/10.1016/0550-3213(84)90015-4}{\emph{Nucl. Phys. B}
  {\bfseries 237} (1984) 59}.

\bibitem{Kocic:1985uq}
A.~Kocic, \emph{{Chiral Symmetry Restoration at Finite Densities in Coulomb
  Gauge {QCD}}}, \href{https://doi.org/10.1103/PhysRevD.33.1785}{\emph{Phys.
  Rev. D} {\bfseries 33} (1986) 1785}.

\bibitem{Hirata:1989qp}
M.~Hirata, \emph{{Composite Meson Quark Interactions Under the Condition of
  Dynamical Breaking of Chiral Symmetry}},
  \href{https://doi.org/10.1103/PhysRevD.39.1425}{\emph{Phys. Rev. D}
  {\bfseries 39} (1989) 1425}.

\bibitem{Alkofer:1989vr}
R.~Alkofer, P.A.~Amundsen and K.~Langfeld, \emph{{Chiral Symmetry Breaking and
  Pion Properties at Finite Temperatures}},
  \href{https://doi.org/10.1007/BF01555857}{\emph{Z. Phys. C} {\bfseries 42}
  (1989) 199}.

\bibitem{Schmidt:1995gea}
S.M.~Schmidt, D.~Blaschke and Y.L.~Kalinovsky, \emph{{Low-energy theorems in a
  nonlocal chiral quark model at finite temperature}},
  \href{https://doi.org/10.1007/BF01556375}{\emph{Z. Phys. C} {\bfseries 66}
  (1995) 485}.

\bibitem{Reinhardt:2017pyr}
H.~Reinhardt, G.~Burgio, D.~Campagnari, E.~Ebadati, J.~Heffner, M.~Quandt
  et~al., \emph{{Hamiltonian approach to QCD in Coulomb gauge - a survey of
  recent results}}, \href{https://doi.org/10.1155/2018/2312498}{\emph{Adv. High
  Energy Phys.} {\bfseries 2018} (2018) 2312498}
  [\href{https://arxiv.org/abs/1706.02702}{{\ttfamily 1706.02702}}].

\bibitem{Quandt:2018bbu}
M.~Quandt, E.~Ebadati, H.~Reinhardt and P.~Vastag, \emph{{Chiral symmetry
  restoration at finite temperature within the Hamiltonian approach to QCD in
  Coulomb gauge}},
  \href{https://doi.org/10.1103/PhysRevD.98.034012}{\emph{Phys. Rev. D}
  {\bfseries 98} (2018) 034012}
  [\href{https://arxiv.org/abs/1806.04493}{{\ttfamily 1806.04493}}].

\bibitem{conf}
P.M.~Lo and E.S.~Swanson, \emph{{Confinement Models at Finite Temperature and
  Density}}, \href{https://doi.org/10.1103/PhysRevD.81.034030}{\emph{Phys. Rev.
  D} {\bfseries 81} (2010) 034030}
  [\href{https://arxiv.org/abs/0908.4099}{{\ttfamily 0908.4099}}].

\bibitem{Klevansky1992}
S.P.~Klevansky, \emph{{The Nambu—Jona-Lasinio model of quantum
  chromodynamics}}, \href{https://doi.org/10.1103/RevModPhys.64.649}{\emph{Rev.
  Mod. Phys.} {\bfseries 64} (1992) 649}.

\bibitem{Kapusta:2006pm}
J.I.~Kapusta and C.~Gale, \emph{{Finite-temperature field theory: Principles
  and applications}}, Cambridge Monographs on Mathematical Physics, Cambridge
  University Press (2011),
  \href{https://doi.org/10.1017/CBO9780511535130}{10.1017/CBO9780511535130}.

\bibitem{Andersen:2014xxa}
J.O.~Andersen, W.R.~Naylor and A.~Tranberg, \emph{{Phase diagram of QCD in a
  magnetic field: A review}},
  \href{https://doi.org/10.1103/RevModPhys.88.025001}{\emph{Rev. Mod. Phys.}
  {\bfseries 88} (2016) 025001}
  [\href{https://arxiv.org/abs/1411.7176}{{\ttfamily 1411.7176}}].

\bibitem{Miransky:2015ava}
V.A.~Miransky and I.A.~Shovkovy, \emph{{Quantum field theory in a magnetic
  field: From quantum chromodynamics to graphene and Dirac semimetals}},
  \href{https://doi.org/10.1016/j.physrep.2015.02.003}{\emph{Phys. Rept.}
  {\bfseries 576} (2015) 1} [\href{https://arxiv.org/abs/1503.00732}{{\ttfamily
  1503.00732}}].

\bibitem{debye}
A.~Bandyopadhyay, B.~Karmakar, N.~Haque and M.G.~Mustafa, \emph{Pressure of a
  weakly magnetized hot and dense deconfined qcd matter in one-loop
  hard-thermal-loop perturbation theory},
  \href{https://doi.org/10.1103/PhysRevD.100.034031}{\emph{Phys. Rev. D}
  {\bfseries 100} (2019) 034031}.

\bibitem{Fukushima:2003fw}
K.~Fukushima, \emph{{Chiral effective model with the Polyakov loop}},
  \href{https://doi.org/10.1016/j.physletb.2004.04.027}{\emph{Phys. Lett. B}
  {\bfseries 591} (2004) 277}
  [\href{https://arxiv.org/abs/hep-ph/0310121}{{\ttfamily hep-ph/0310121}}].

\bibitem{pnjl}
C.~Ratti, M.A.~Thaler and W.~Weise, \emph{Phases of qcd: Lattice thermodynamics
  and a field theoretical model},
  \href{https://doi.org/10.1103/PhysRevD.73.014019}{\emph{Phys. Rev. D}
  {\bfseries 73} (2006) 014019}
  [\href{https://arxiv.org/abs/hep-ph/0506234}{{\ttfamily hep-ph/0506234}}].

\bibitem{Sasaki:2006ww}
C.~Sasaki, B.~Friman and K.~Redlich, \emph{{Susceptibilities and the Phase
  Structure of a Chiral Model with Polyakov Loops}},
  \href{https://doi.org/10.1103/PhysRevD.75.074013}{\emph{Phys. Rev. D}
  {\bfseries 75} (2007) 074013}
  [\href{https://arxiv.org/abs/hep-ph/0611147}{{\ttfamily hep-ph/0611147}}].

\bibitem{Fukushima:2017csk}
K.~Fukushima and V.~Skokov, \emph{{Polyakov loop modeling for hot QCD}},
  \href{https://doi.org/10.1016/j.ppnp.2017.05.002}{\emph{Prog. Part. Nucl.
  Phys.} {\bfseries 96} (2017) 154}
  [\href{https://arxiv.org/abs/1705.00718}{{\ttfamily 1705.00718}}].

\bibitem{Td1}
{\scshape HotQCD} collaboration, \emph{{Chiral crossover in QCD at zero and
  non-zero chemical potentials}},
  \href{https://doi.org/10.1016/j.physletb.2019.05.013}{\emph{Phys. Lett. B}
  {\bfseries 795} (2019) 15}
  [\href{https://arxiv.org/abs/1812.08235}{{\ttfamily 1812.08235}}].

\bibitem{Td2}
S.~Borsanyi, Z.~Fodor, J.N.~Guenther, R.~Kara, S.D.~Katz, P.~Parotto et~al.,
  \emph{{The QCD crossover at finite chemical potential from lattice
  simulations}},
  \href{https://doi.org/10.1103/PhysRevLett.125.052001}{\emph{Phys. Rev. Lett.}
  {\bfseries 125} (2020) 052001}
  [\href{https://arxiv.org/abs/2002.02821}{{\ttfamily 2002.02821}}].

\bibitem{shift1}
B.-J.~Schaefer, J.M.~Pawlowski and J.~Wambach, \emph{{The Phase Structure of
  the Polyakov--Quark-Meson Model}},
  \href{https://doi.org/10.1103/PhysRevD.76.074023}{\emph{Phys. Rev. D}
  {\bfseries 76} (2007) 074023}
  [\href{https://arxiv.org/abs/0704.3234}{{\ttfamily 0704.3234}}].

\bibitem{shift2}
J.~Braun, H.~Gies and J.M.~Pawlowski, \emph{{Quark Confinement from Color
  Confinement}},
  \href{https://doi.org/10.1016/j.physletb.2010.01.009}{\emph{Phys. Lett. B}
  {\bfseries 684} (2010) 262}
  [\href{https://arxiv.org/abs/0708.2413}{{\ttfamily 0708.2413}}].

\bibitem{mediumG1}
G.~Endr\H{o}di and G.~Mark\'o, \emph{{Magnetized baryons and the QCD phase
  diagram: NJL model meets the lattice}},
  \href{https://doi.org/10.1007/JHEP08(2019)036}{\emph{JHEP} {\bfseries 08}
  (2019) 036} [\href{https://arxiv.org/abs/1905.02103}{{\ttfamily
  1905.02103}}].

\bibitem{Skokov:2010wb}
V.~Skokov, B.~Stokic, B.~Friman and K.~Redlich, \emph{{Meson fluctuations and
  thermodynamics of the Polyakov loop extended quark-meson model}},
  \href{https://doi.org/10.1103/PhysRevC.82.015206}{\emph{Phys. Rev. C}
  {\bfseries 82} (2010) 015206}
  [\href{https://arxiv.org/abs/1004.2665}{{\ttfamily 1004.2665}}].

\bibitem{Skokov:2010uh}
V.~Skokov, B.~Friman and K.~Redlich, \emph{{Quark number fluctuations in the
  Polyakov loop-extended quark-meson model at finite baryon density}},
  \href{https://doi.org/10.1103/PhysRevC.83.054904}{\emph{Phys. Rev. C}
  {\bfseries 83} (2011) 054904}
  [\href{https://arxiv.org/abs/1008.4570}{{\ttfamily 1008.4570}}].

\bibitem{su3pot}
P.M.~Lo, K.~Redlich and C.~Sasaki, \emph{{Fluctuations of the order parameter
  in an $SU(N_c)$ effective model}},
  \href{https://doi.org/10.1103/PhysRevD.103.074026}{\emph{Phys. Rev. D}
  {\bfseries 103} (2021) 074026}
  [\href{https://arxiv.org/abs/2101.12663}{{\ttfamily 2101.12663}}].

\bibitem{regulators}
S.S.~Avancini, R.L.S.~Farias, N.N.~Scoccola and W.R.~Tavares, \emph{Njl-type
  models in the presence of intense magnetic fields: The role of the
  regularization prescription},
  \href{https://doi.org/10.1103/PhysRevD.99.116002}{\emph{Phys. Rev. D}
  {\bfseries 99} (2019) 116002}.

\bibitem{Lo:2013lca}
P.M.~Lo and E.S.~Swanson, \emph{{QED3 at Finite Temperature and Density}},
  \href{https://doi.org/10.1103/PhysRevD.89.025015}{\emph{Phys. Rev. D}
  {\bfseries 89} (2014) 025015}
  [\href{https://arxiv.org/abs/1307.5834}{{\ttfamily 1307.5834}}].

\bibitem{Lo:2013hla}
P.M.~Lo, B.~Friman, O.~Kaczmarek, K.~Redlich and C.~Sasaki, \emph{{Polyakov
  loop fluctuations in SU(3) lattice gauge theory and an effective gluon
  potential}}, \href{https://doi.org/10.1103/PhysRevD.88.074502}{\emph{Phys.
  Rev. D} {\bfseries 88} (2013) 074502}
  [\href{https://arxiv.org/abs/1307.5958}{{\ttfamily 1307.5958}}].

\bibitem{sus}
P.M.~Lo, B.~Friman, O.~Kaczmarek, K.~Redlich and C.~Sasaki, \emph{{Probing
  Deconfinement with Polyakov Loop Susceptibilities}},
  \href{https://doi.org/10.1103/PhysRevD.88.014506}{\emph{Phys. Rev. D}
  {\bfseries 88} (2013) 014506}
  [\href{https://arxiv.org/abs/1306.5094}{{\ttfamily 1306.5094}}].

\bibitem{Bali:2011qj}
G.S.~Bali, F.~Bruckmann, G.~Endrodi, Z.~Fodor, S.D.~Katz, S.~Krieg et~al.,
  \emph{{The QCD phase diagram for external magnetic fields}},
  \href{https://doi.org/10.1007/JHEP02(2012)044}{\emph{JHEP} {\bfseries 02}
  (2012) 044} [\href{https://arxiv.org/abs/1111.4956}{{\ttfamily 1111.4956}}].

\bibitem{Bruckmann:2013oba}
F.~Bruckmann, G.~Endrodi and T.G.~Kovacs, \emph{{Inverse magnetic catalysis and
  the Polyakov loop}},
  \href{https://doi.org/10.1007/JHEP04(2013)112}{\emph{JHEP} {\bfseries 04}
  (2013) 112} [\href{https://arxiv.org/abs/1303.3972}{{\ttfamily 1303.3972}}].

\bibitem{Boomsma:2009yk}
J.K.~Boomsma and D.~Boer, \emph{{The Influence of strong magnetic fields and
  instantons on the phase structure of the two-flavor NJL model}},
  \href{https://doi.org/10.1103/PhysRevD.81.074005}{\emph{Phys. Rev. D}
  {\bfseries 81} (2010) 074005}
  [\href{https://arxiv.org/abs/0911.2164}{{\ttfamily 0911.2164}}].

\bibitem{Lo:2020ptj}
P.M.~Lo, M.~Szyma\'nski, C.~Sasaki and K.~Redlich, \emph{{Deconfinement in the
  presence of a strong magnetic field}},
  \href{https://doi.org/10.1103/PhysRevD.102.034024}{\emph{Phys. Rev. D}
  {\bfseries 102} (2020) 034024}
  [\href{https://arxiv.org/abs/2004.04138}{{\ttfamily 2004.04138}}].

\bibitem{mediumG2}
S.S.~Avancini, R.L.S.~Farias, M.~Benghi~Pinto, W.R.~Tavares and V.S.~Tim\'oteo,
  \emph{{$\pi_0$ pole mass calculation in a strong magnetic field and lattice
  constraints}},
  \href{https://doi.org/10.1016/j.physletb.2017.02.002}{\emph{Phys. Lett. B}
  {\bfseries 767} (2017) 247}
  [\href{https://arxiv.org/abs/1606.05754}{{\ttfamily 1606.05754}}].

\bibitem{eric_cgauge}
A.P.~Szczepaniak and E.S.~Swanson, \emph{{Coulomb gauge QCD, confinement, and
  the constituent representation}},
  \href{https://doi.org/10.1103/PhysRevD.65.025012}{\emph{Phys. Rev. D}
  {\bfseries 65} (2001) 025012}
  [\href{https://arxiv.org/abs/hep-ph/0107078}{{\ttfamily hep-ph/0107078}}].

\end{thebibliography}\endgroup


\end{document}